\documentclass[final]{svjour2}

\usepackage{graphicx}
\usepackage{rotating}
\usepackage{amssymb}
\usepackage{mathptmx}
\usepackage[numbers]{natbib}
\usepackage[squaren,thinqspace]{SIunits}

\makeatletter
\journalname{Journal of Low Temperature Physics}

\bibpunct{}{}{,}{s}{}{,}

\begin{document}

\title{Multiplexed readout of MMC detector arrays using non-hysteretic rf-SQUIDs}

\author{S. Kempf \and M. Wegner \and L. Gastaldo \and A. Fleischmann \and C. Enss}

\institute{S. Kempf \and M. Wegner \and L. Gastaldo \and A. Fleischmann \and C. Enss \at
           Kirchhoff-Institute for Physics, Heidelberg University, Im Neuenheimer Feld 227, 69120 Heidelberg, Germany \\
           \email{sebastian.kempf@kip.uni-heidelberg.de}
}

\date{15.07.2013}

\maketitle


\begin{abstract}

Metallic magnetic calorimeters (MMCs) are widely used for various experiments in fields ranging from atomic and nuclear physics to x-ray spectroscopy, laboratory astrophysics or material science. Whereas in previous experiments single pixel detectors or small arrays have been used, for future applications large arrays are needed. Therefore, suitable multiplexing techniques for MMC arrays are currently under development. A promising approach for the readout of large arrays is the microwave SQUID multiplexer that operates in the frequency domain and that employs non-hysteretic rf-SQUIDs to transduce the detector signals into a frequency shift of high $Q$ resonators which can be monitored by using standard microwave measurement techniques. In this paper we discuss the design and the expected performance of a recently developed and fabricated 64 pixel detector array with integrated microwave SQUID multiplexer. First experimental data were obtained characterizing dc-SQUIDs with virtually identical washer design.

\keywords{microwave SQUID multiplexer, metallic magnetic calorimeters, integrated array readout}

\end{abstract}


\section{Introduction}

The very high energy resolution, the intrinsically dissipationless nature of operation, the inherent fast pulse rise time and the excellent linearity are only a few out of many properties that make metallic magnetic calorimeters \cite{Fle05} (MMCs) particulary attractive candidates for a variety of applications requiring energy dispersive particle detectors with high time resolution. State-of-the-art MMCs, for example, being optimized for soft x-ray spectroscopy have shown an energy resolution of $\Delta E_\mathrm{FWHM} = \unit{2}{\electronvolt}$ at a photon energy of \unit{6}{\kilo\electronvolt} and a rise formation time of \unit{90}{\nano\second} \cite{Pie12}. Due to the consequent use of microfabrication techniques for device fabrication, detector arrays consisting of tens or hundreds of virtually identical MMCs can be routinely produced. However, at present, only small detector arrays are used since multiplexing of MMCs without significant degradation of the single pixel performance such as the energy resolution or the rise time has not yet been established.

Very recently, time domain SQUID multiplexing (TDM) of small arrays was successfully demonstrated\cite{Por13}. The obtained results are promising and show that small arrays can be read out without a significant degradation of the energy resolution. However, TDM of arrays consisting of several hundred or thousand detectors is difficult to imagine since the apparent energy sensitivity $\epsilon_\mathrm{s}$ of the SQUIDs increases with the number of detectors due to wideband SQUID noise aliasing that results from the required large open-loop bandwidth of the SQUIDs combined with the less often sampling of the detector signals\cite{Irw09}.
In addition, the bandwidth of each detector has to be limited in order to avoid detector noise aliasing\cite{Irw09}. This prevents the use of TDM for applications requiring a short response time. In contrast, a microwave SQUID multiplexer\cite{Irw04, Mat08} is well suited for the readout of large arrays since $\epsilon_\mathrm{s}$ is independent of the number of detectors and the bandwidth per pixel needs not to be limited to avoid detector noise aliasing.

In this paper we discuss the design of a recently developed and fabricated 64 pixel detector array with integrated microwave SQUID multiplexer. This array was produced to test the suitability of this readout technique and will be used in a first small-scale ECHo experiment\cite{Gas13}.


\section{Principle of the microwave SQUID multiplexer}

\begin{figure}
	\begin{center}
		\includegraphics[width=0.7\linewidth,keepaspectratio] {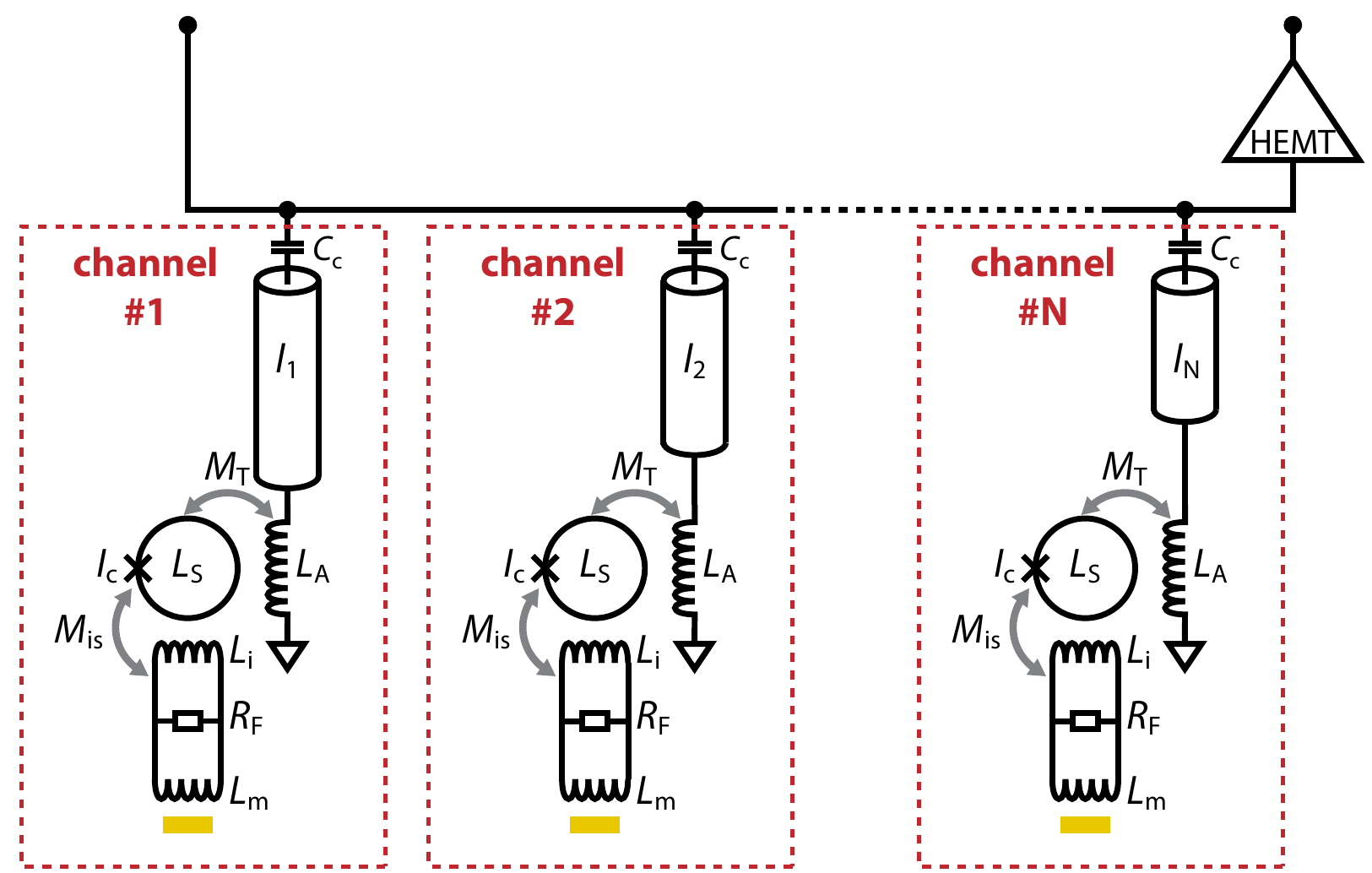}
	\end{center}
\caption{(Color online) Schematic of a detector array read out by using a microwave SQUID multiplexer that is based on
	dissipationless, non-hysteretic rf-SQUIDs.}
\label{fig:SQMUX}
\end{figure}

Figure~\ref{fig:SQMUX} shows a schematic of a detector array read out using a microwave SQUID multiplexer that is based on dissipationless, non-hysteretic rf-SQUIDs. Each detector is inductively coupled to an unshunted single-junction SQUID with loop inductance $L_\mathrm{s}$ and critical current $I_\mathrm{c}$. For $\beta_\mathrm{L} \equiv 2\pi L_\mathrm{s} I_\mathrm{c}/\Phi_0 < 1$ with $\Phi_0$ denoting the magnetic flux quantum, the SQUID is non-hysteretic, i.e. it behaves purely reactive, and can be modelled as a non-linear inductor $L(\Phi)$ whose value depends on the magnetic flux $\Phi$ threading the SQUID. In order to read out the change of $L(\Phi)$ that is caused by a magnetic flux change $\Phi$ associated with a detector signal, the SQUID is inductively coupled to the load inductor $L_\mathrm{A}$ terminating the associated high $Q$ superconducting transmission line resonator with unique resonance frequency in the \giga\hertz\ range that is defined by the resonator length. Due to the mutual interaction with the SQUID, the actual value $L_\mathrm{A}(\Phi)$ of the load inductor is also flux dependent and therefore the circuit's resonance frequency gets shifted as the magnetic flux $\Phi$ changes. Assuming the circuit is excited with a fixed microwave signal, this frequency shift can be monitored as a change of amplitude or phase by using a homodyne detection technique. Furthermore, by capacitively coupling many of those circuits to a common transmission line, injecting a microwave frequency comb driving each circuit at resonance and monitoring the amplitude or phase of each resonator, it is possible to measure the signals of a quite large number of detectors simultaneously. In particular, just one HEMT amplifier and two coaxial cables are required for the readout of some hundreds of detectors.


\section{Design of the 64 pixel detector array with integrated SQUID readout}

\begin{figure}
	\begin{center}
		\includegraphics[width=1.0\linewidth,keepaspectratio] {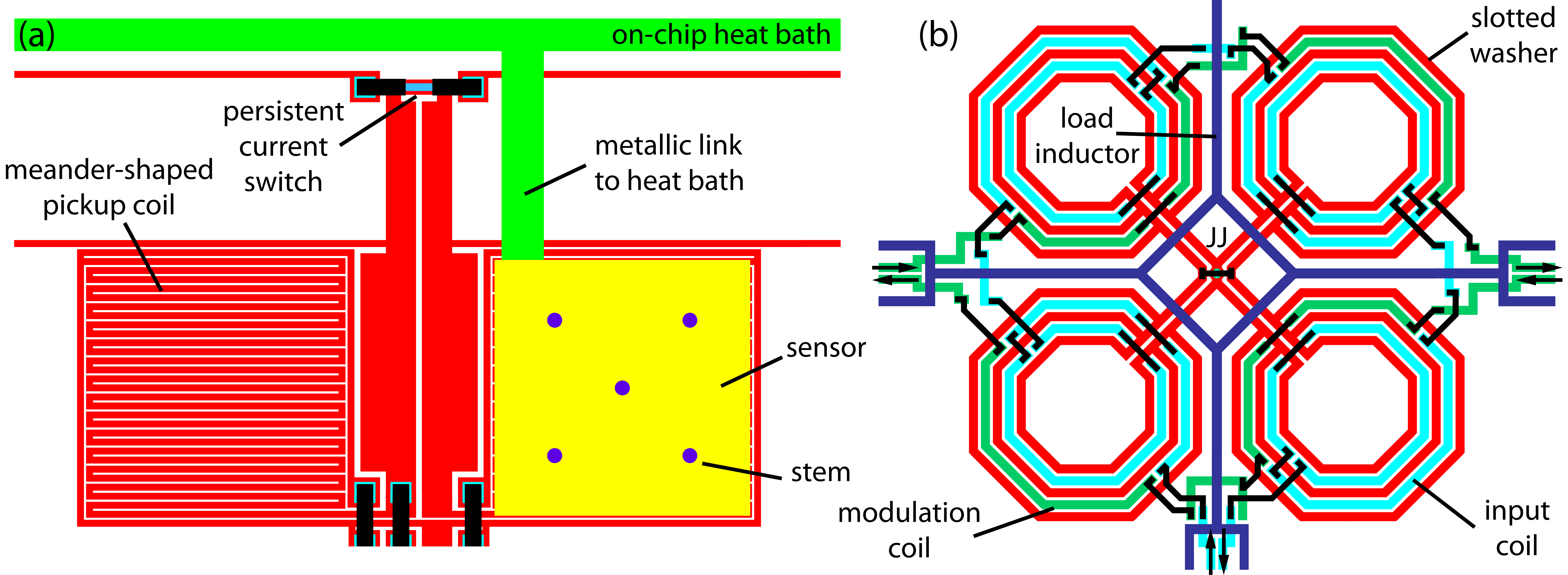}
	\end{center}
\caption{(Color online) ({\it a}) Schematic of the developed two-pixel detector. The sensor, the stems as well as the
	metallic link to the heat bath of the second pixel are not shown for clarity. ({\it b}) Schematic of the developed rf-SQUID.
	The upper part of the load inductor is connected to the microwave resonator while the other ends are connected to ground. The
	Josephson junction (JJ) is located in the middle of the SQUID.
}
\label{fig:design_pictures}
\end{figure}

The detector array being discussed within this paper consists of 32 two-pixel detectors that are depicted in figure~\ref{fig:design_pictures}a) and that are arranged in a linear $2 \times 16$ configuration. Every detector features two electroplated Au absorbers, each having an area of $\unit{170}{\micro\meter \times \unit{170}{\micro\meter}}$ and a thickness of $2\times\unit{5}{\micro\meter}$. In between the two \unit{5}{\micro\meter} thick Au layers, $^{163}$Ho can be deposited within a slightly reduced area, e.g. by means of ion implantation, resulting in a $4\pi$ geometry with a quantum efficiency of more than $\unit{99.9}{\%}$ for particles emitted during the decay of $^{163}$Ho. Each absorber is connected via 5 stems with a diameter of \unit{10}{\micro\meter} to an underlying planar Au:Er$_{\unit{300}{ppm}}$ temperature sensor having an area of $\unit{170}{\micro\meter} \times \unit{170}{\micro\meter}$ and a thickness of \unit{1.35}{\micro\meter}. Since the effective contact area between sensor and absorber is only about \unit{1.4}{\%}, a loss of athermal phonons to the solid substrate is greatly reduced\cite{Fle09}. Underneath both sensors, superconducting meander-shaped pickup coils with a linewidth of \unit{3}{\micro\meter} and a pitch of \unit{6}{\micro\meter} are placed. They are gradiometrically connected in parallel with the input coil of the current-sensing rf-SQUID reading out the detector. This circuitry allows to store a persistent current to create the bias magnetic field being required to magnetize the sensor and to read out two pixels using a single SQUID. To adjust the signal decay time to \unit{1}{\milli\second}, each sensor is connected via a Au link with a low temperature resistance of about \unit{1}{\ohm} to an on-chip heat bath made of electroplated Au.

The inductance $L_\mathrm{i}$ of the input coil of the SQUID reading out the detector has a value of \unit{1.3}{\nano\henry} and is matched to the inductance $L_\mathrm{m}$ of the meander-shaped pickup coils, thus ensuring maximum flux coupling between detector and SQUID. Due to the resistor $R_\mathrm{F} = \unit{2}{\ohm}$ (see figure~\ref{fig:SQMUX}) that is placed in parallel to the input coil, the frequency response of the superconducting flux transformer formed by $L_\mathrm{i}$ and $L_\mathrm{m}$ has a low-pass characteristic with a cutoff frequency of about \unit{500}{\mega\hertz}. This prevents microwave power to leak into the detector.

The rf-SQUID that is depicted in figure~\ref{fig:design_pictures}b) is a second order parallel gradiometer. It is formed by four slotted octagonal washers, each of it having a nominal inductance of \unit{200}{\pico\henry}, that are connected in parallel, hence resulting in a total SQUID inductance of $L_\mathrm{s} = \unit{50}{\pico\henry}$. The critical current of the Josephson junction is designed to be \unit{5}{\micro\ampere} translating into a hysteresis parameter of $\beta_\mathrm{L} = 0.76$. The slotted washer design is chosen to reduce parasitic capacitive coupling effects between the SQUID and the input coil. The SQUID is furthermore equipped with a common flux modulation coil that runs through all SQUIDs to allow for a simultaneous flux biasing or flux ramp modulation\cite{Mat12} of all SQUIDs.

The coplanar transmission line resonators have a resonance frequency $f_\mathrm{r}$ between \unit{4}{\giga\hertz} and \unit{6}{\giga\hertz} which is set by their geometrical length. The coupling capacitance $C_\mathrm{c}$ (see figure~\ref{fig:SQMUX}) is in each case chosen to achieve a loaded quality factor $Q_\mathrm{l} = 5000$ and is implemented by running a part of each resonator in parallel to the common transmission line having a characteristic impedance of $\unit{50}{\ohm}$. The bandwidth of each resonator is large enough to resolve the expected fast signal rise time of a detector which is expected to be in the order of \unit{500}{\nano\second}. The mutual inductance between the SQUID and the load inductance $L_\mathrm{A}$ is about \unit{3}{\pico\henry} which leads to a peak-to-peak resonance frequency shift that equals twice the bandwidth of the resonator. This ensures high SQUID gain and stable resonator operation without the occurence of bifurcations\cite{Sid05} at relatively high microwave powers that are required to strongly drive the SQUID.

Taking into account all design parameters and assuming that the SQUID is driven hard, i.e. the antinode current $I_\mathrm{A}$ that oscillate the magnetic flux inside the SQUID is set to $I_\mathrm{A} = \Phi_0/2\pi M_\mathrm{T}$, the expected energy sensitivity of the SQUIDs is around \unit{600}{\hbar} depending on the actual resonance frequency of the related resonator. This translates into an expected energy resolution of about \unit{5}{\electronvolt} of the detectors.


\section{Characterization dc-SQUIDs with virtually identical washer design}

The characterization of the rf-SQUIDs that are present in the microwave SQUID multiplexer can not be directly done. In order to quantify important paramaters such as the critical current $I_0$ or the capacitance $C_\mathrm{J}$ of the Josephson junction or the SQUID inductance $L_\mathrm{s}$, we also produced dc-SQUIDs with virtually identical washer design that allows to survey some important parameters at \unit{4.2}{\kelvin}. Both junctions of the dc-SQUIDs were shunted with $R_\mathrm{N} = \unit{4}{\ohm}$ and their critical current $I_0$ were tripled compared to the rf-SQUIDs. But, due to a design flaw which was recognized not before the characterziation, the SQUID is a first order gradiometer with $L_\mathrm{s} = \unit{200}{\pico\henry}$ instead of a second order gradiometer with $L_\mathrm{s} = \unit{50}{\pico\henry}$. This makes the dc-SQUIDs slightly hysteretic since $\beta \equiv 2 L_\mathrm{s} I_0 /\Phi_0 > 1$. However, a partial SQUID characterization could still be performed.

\begin{figure}
	\begin{center}
		\includegraphics[width=0.9\linewidth,keepaspectratio] {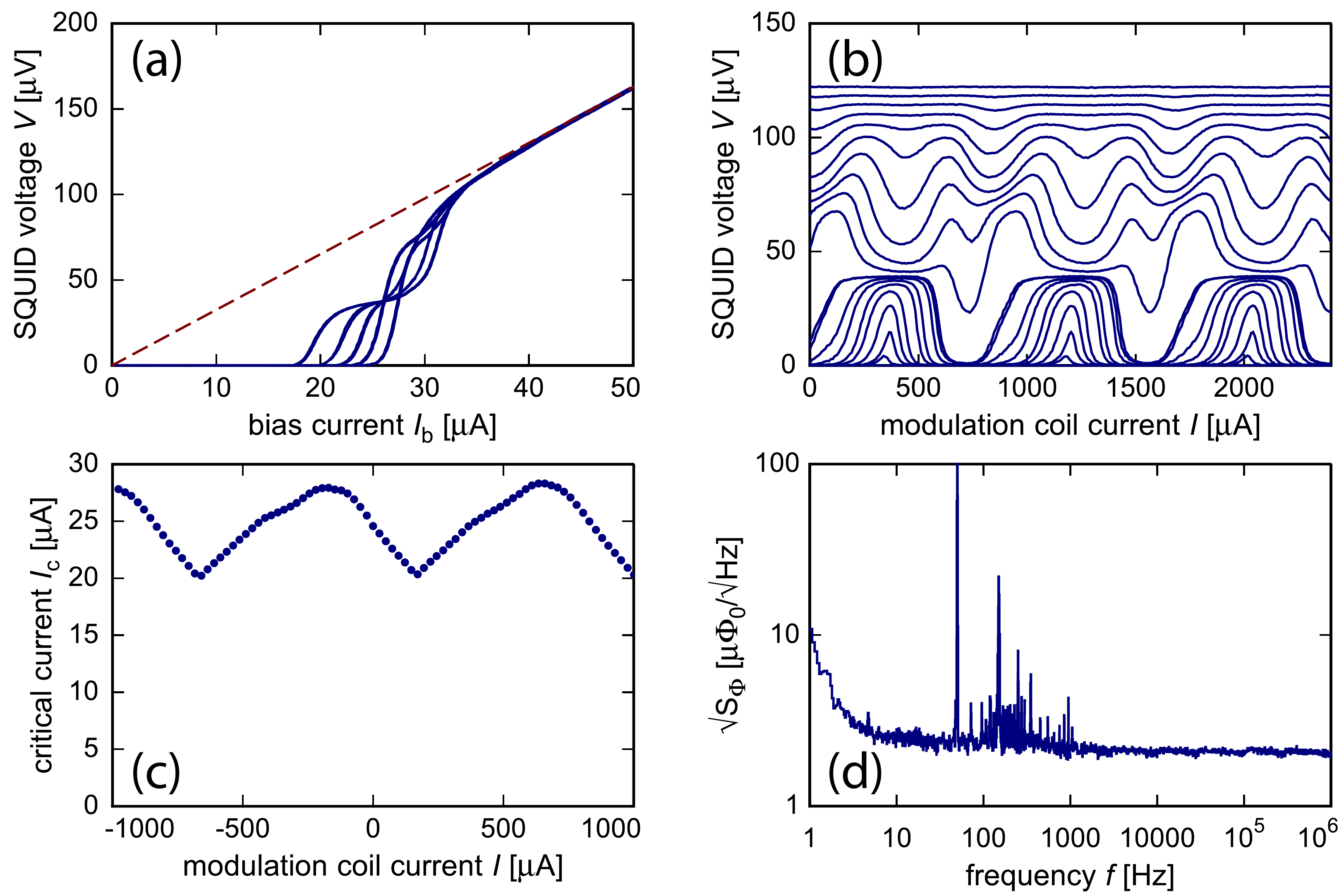}
	\end{center}
\caption{(Color online) Representative ({\it a}) $V$-$I$, ({\it b}) $V$-$\Phi$, ({\it c}) $I_\mathrm{c}$-$\Phi$ characteristics 
	and ({\it d}) measured magnetic flux noise of a dc-SQUID being virtually identical to the developed rf-SQUIDs.}
\label{fig:dc_SQUID}
\end{figure}

Figure~\ref{fig:dc_SQUID}a), b) and c) show representative $V$-$I$, $V$-$\Phi$ and $I_\mathrm{c}$-$\Phi$ characteristics of one of our dc-SQUIDs which were measured at \unit{4.2}{\kelvin}. A linear fit to the resistive branch of the $V$-$I$ characteristic reveals a normal state resistance of $R_\mathrm{N} = \unit{6.4}{\ohm}$ which is about \unit{60}{\%} larger than our design value and is most likely caused by an insufficient parameter control during lithography. However, this affects the performance of the SQUID multiplexer just marginally since the junctions are unshunted. Solely the cutoff frequency of the low-pass filter within the input circuit is increased to about \unit{1}{\giga\hertz} which is however still sufficient to prevent microwave power to leak into the detectors. The $I_\mathrm{c}$-$\Phi$ characteristic shows a $I_\mathrm{c}$ modulation of $\Delta I_\mathrm{c}/I_\mathrm{c}^{max} = 0.27$ with $I_\mathrm{c}^\mathrm{max} = \unit{27.8}{\micro\ampere}$ and is slightly asymmetric which we attribute to an asymmetry in the shunt resistors, again most likely due to an insufficient parameter control. According to {\it Tesche and Clarke}\cite{Tes77} we get a SQUID inductance of $L_\mathrm{s} = \unit{186}{\pico\henry}$ taking into account $\beta$ as determined from the measured value of $\Delta I_\mathrm{c}/I_\mathrm{c}^{max}$. Both, the junction critical current $I_0$ and the inductance $L_\mathrm{w} =\unit{186}{\pico\henry}$ of a single slotted washer, hence agree within \unit{10}{\%} with our design values and anticipate a succesful operation of the microwave SQUID multiplexer.

Since the washers are not resistively damped, the fundamental SQUID resonance appears at a voltage of \unit{37}{\micro\volt}. This value reveals a junction capacitance of $C_\mathrm{J} = \unit{0.85}{\pico\farad}$ which is about \unit{70}{\%} smaller than the calculated value taking into accout the critical current density of $j_\mathrm{c} = \unit{40}{\ampere\per\centi\meter\squared}$ and the junction area\cite{Mae95}. The reason for this is not yet clear, but will be investigated more in detail in near future.

Figure~\ref{fig:dc_SQUID}d) shows the measured magnetic flux noise of one of our SQUIDs. The white noise level of this device is $\sqrt{S_{\Phi, \mathrm{w}}} = \unit{2.1}{\micro\Phi_0/\sqrt\hertz}$ and includes a contribution of $\sqrt{S_{\Phi, \mathrm{el}}} = \unit{0.9}{\micro\Phi_0/\sqrt\hertz}$ of the direct readout electronics. The intrinsic flux noise is $\sqrt{S_{\Phi, \mathrm{int}}} = \unit{1.9}{\micro\Phi_0/\sqrt\hertz}$ and is about \unit{35}{\%} larger than the calculated value assuming that $\sqrt{S_{\Phi, \mathrm{int}}}$ is soley determined by the current noise of the shunt resistors. But, since $\beta > 1$ and $\beta_c \equiv 2\pi C_\mathrm{J} R_\mathrm{N}^2 I_0 / \Phi_0 > 1$, resonances are very likely to occur that might not be observed in the SQUID characteristics but appear as an increased white noise level. At low frequencies a $1/f$ like noise contribution with a corner frequency of about \unit{3}{\hertz} is observed. It can be attributed rather to external pertubations due to the magnetically unshielded SQUID operation than to critical current fluctuations or magnetic impurities in the vicinity of the SQUID.


\section{Conclusions}

We have summarized the design and the expected performance of a recently developed and fabricated 64 pixel detector array with integrated microwave SQUID multiplexer that was produced to test the suitability of this readout technique and that will be used in a first small-scale ECHo experiment. The characterization of dc-SQUIDs with virtually identical washer design revealed that the crucial SQUID parameters such as the critical current of the Josehson junctions or the washer inductance are close to the design values and anticipates a successful operation of the SQUID multiplexer which will be tested in near future.


\begin{acknowledgements}
We would like to thank J. Beyer, G.C. Hilton, J.A.B. Mates and K.D. Irwin for many fruitful discussions. We are also very grateful to T. Wolf and V. Schultheiss for technical support during device fabrication. This work was supported by the BMBF grant 06 HD 9118I, the GSI R\&D grant HDEnss, and the European Community Research Infrastructures under the FP7 Capacities Specific Programme, MICROKELVIN project number 228464.
\end{acknowledgements}




\begin{thebibliography}{99}

\bibitem{Fle05}
A. Fleischmann et al., in {\em Cryogenic Particle Detection} ed. by C. Enss, Topics in Applied Physics, vol. 99, Springer (2005)
\bibitem{Pie12}
C. Pies et al., J. Low Temp. Phys. {\bf 167}, 269--279 (2012)
\bibitem{Por13}
J.-P. Porst et al., IEEE Trans. Appl. Supercond. {\bf 23}, 2500905 (2013)
\bibitem{Irw09}
K.D. Irwin, AIP Conf. Proc. {\bf 1185}, 229--236 (2009)
\bibitem{Irw04}
K.D. Irwin et al., Appl. Phys. Lett. {\bf 85}, 2107-2109 (2004)
\bibitem{Mat08}
J.A.B. Mates et al., Appl. Phys. Lett. {\bf 92}, 023513 (2008)
\bibitem{Gas13}
L. Gastaldo, these proceedings (2013)
\bibitem{Fle09}
A. Fleischmann et al., AIP Conf. Proc. {\bf 1185}, 571--578 (2009)
\bibitem{Mat12}
J.A.B. Mates et al., J. Low Temp. Phys. {\bf 167}, 707--712 (2012)
\bibitem{Sid05}
I. Siddiqi et al., Phys. Rev. Lett. {\bf 94}, 27005 (2005)
\bibitem{Tes77}
C.D. Tesche et al., J. Low Temp. Phys. {\bf 29}, 301--331 (1977)
\bibitem{Mae95}
M. Maezawa et al., Appl. Phys. Lett., {\bf 66}, 2134--2136 (1995)

\end{thebibliography}
\end{document}